\documentclass[twocolumn, showpacs,aps, prl, amsmath,amssymb,superscriptaddress, floatfix]{revtex4}
\usepackage{graphicx}
\newcommand{\bb}{\begin{equation}}
\newcommand{\en}{\end{equation}}
 
\begin{document}
 \title{Equilibrium Bundle Size of Rodlike Polyelectrolytes with Counterion-Induced Attractive Interactions}
 
 \author{Mark L. Henle}
 \affiliation{Department of Physics, University of California, Santa Barbara, California 93106}
 \author{Philip A. Pincus}
 \affiliation{Department of Physics, University of California, Santa Barbara, California 93106}
\affiliation{Department of Materials, University of California, Santa Barbara, California 93106}
\affiliation{Program in Biomolecular Science and Engineering, University of California, Santa Barbara, California 93106}
\affiliation{Department of Physics, Korea Advanced Institute of Science and Technology, Daejon 305-701, Republic of Korea}

\date{\today}
\begin{abstract}

Multivalent counterions can induce an effective attraction between like-charged rodlike polyelectrolytes, leading to the formation of polelectrolyte bundles.  In this paper, we calculate the equilibrium bundle size using a simple model in which the attraction between polyelectrolytes (assumed to be pairwise additive) is treated phenomenologically.  If the counterions are point-like, they almost completely neutralize the charge of the bundle, and the equilibrium bundle size diverges. When the counterions are large, however, steric and short-range electrostatic interactions prevent charge neutralization of the bundle, thus forcing the equilibrium bundle size to be finite.  We also consider the possibility that increasing the number of nearest neighbors for each rod in the bundle frustrates the attractive interaction between the rods.  Such a frustration leads to the formation of finite size bundles as well, even when the counterions are small.  

\end{abstract}

\pacs{61.20.Qg, 61.25.Hq, 82.35.Rs, 82.70-y}
 
 \maketitle
 
The mean-field Poisson-Boltzmann (PB) theory of electrostatic interactions predicts that two identical macromolecules in any salt solution will repel each other \cite{Sader}.  However, the presence of multivalent counterions can actually induce an \emph{attraction} between like-charged polyelectrolytes (PEs).  This has been experimentally observed for several different PEs, including double-stranded DNA \cite{Bloomfield,Hud}, F-actin \cite{Tang1,Angelini}, microtubules \cite{Tang1,Needleman1}, and the fd, M13, and tobacco mosaic viruses \cite{Tang1,Tang2}.  Computer simulations of both homogeneously charged rods \cite{Jensen,Stevens, Deserno1,Guldbrand} and realistic DNA molecules \cite{Guldbrand,Nordenskiold,Allahyarov} unambiguously show that attractive interactions can arise solely from counterion correlations not included in PB theory. Several theories that take these correlations into account -- including perturbative expansions of PB theory \cite{Oosawa,Ha1}, structural-correlation theory \cite{Rouzina,Shklovskii, delaCruz1}, and strong-coupling theory \cite{Netz} -- obtain an attractive interaction between two rods.  It is still a matter of discussion, however, as to which of these theories is the most appropriate description of the correlation-induced attraction seen in experiments and simulations.  Furthermore, it is unknown whether the interactions between multiple rods is pairwise additive or not \cite{Podgornik,Ha2,Shklovskii}.  

Under experimental conditions in which the interaction between PEs is attractive, the PEs typically form dense, ordered bundles of a well-defined size, rather than precipitating into a PE-rich phase \cite{Bloomfield, Hud, Tang1, Angelini, Needleman1, Tang2}. In this paper, we theoretically investigate the thermodynamic stability of these bundles (if bundle growth is not limited thermodynamically, then it must be limited by kinetic barriers \cite{Ha2, delaCruz2, Henle}).  We assume that the attractive interactions are pairwise additive, but do not specify the precise nature of the counterion correlations.  Rather, we simply introduce a phenomenological parameter $\gamma$ to characterize the attractive energy between two PEs in a bundle. 

Consider, then, an aqueous solution of volume $V$ with $N$ identical rodlike PEs of length $L$, radius $a_0$, and a uniform linear charge density $-e \lambda_0$ (the aggregation of flexible PEs has been considered in \cite{delaCruz2}).  We treat the aqueous solution as a uniform dielectric with dielectric constant $\epsilon$, and ignore all image-charge effects (i.e. we assume that the dielectric constant of the rods is $\approx \epsilon$).   Positive monovalent and $q$-valent counterions, as well as negative monovalent co-ions, are present;  the entire system is charge neutral and in chemical equilibrium with a salt bath.  It is well known that a highly charged PE will have a large number of counterions ``condensed'' near its surface \cite{Manning}.  A condensed multivalent ion can become strongly correlated both with ions condensed on the same rod and with ions condensed on nearby rods.  The latter correlations give rise to the effective attraction between rods, which in turn leads to the formation of PE bundles of some size at equilibrium.  For simplicity, we assume that only multivalent ions   can enter inside the bundle (the effects of competitive binding with monovalent ions will be discussed in a future paper \cite{Henle}). Also, we assume that the solution of rods is dilute, i.e. the volume fraction $\phi \equiv N \pi a_0^2 L/V \ll 1$.  As a result, we can employ the cell model, where each bundle and its surrounding ions are enclosed in a Wigner-Seitz (WS) cell, and interactions between cells are ignored.  We work in the long-rod limit ($L \rightarrow \infty$), so that the translational entropy of the bundles is negligible.  Finally,  we assume that the equilibrium distribution of bundle sizes is sharply peaked, so that all bundles in the system have approximately the same size.  Given these assumptions,the free energy can be written as a sum of four terms: 
\bb
\label{eq:freeEnergy}
\beta F = \frac{N}{M}  \left[ \beta F_{ent}+\beta F_{ES} + \beta F_{attr}\right]+N \beta F_{corr}
\en
where $M$ is the number of rods in each bundle and $\beta \equiv 1/k_B T$, $k_B$ being Boltzmann's constant and $T$ the temperature.  $\beta F_{ent}$ includes the entropy and the chemical potential terms for all of the ions in a single WS cell; $\beta F_{ES}$ is the total mean-field electrostatic energy for a WS cell; $\beta F_{attr}$ is the total correlation-induced attractive energy for a single bundle; and $\beta F_{corr}$ is the correlation energy for ions condensed on a single rod.   

To calculate $\beta F_{ent}$, we assume that the ions inside the bundle are uniformly distributed throughout the volume available inside the bundle.  The ions outside the bundle, on the other hand, are treated with a modified Debye-Huckel (DH) theory similar to Manning's counterion condensation theory \cite{Manning}.  It is well known that DH theory breaks down near highly charged surfaces \cite{Schiessel}; in particular, it does not properly account for counterion condensation.  To correct for this, we allow ions to condense into a Stern layer surrounding the bundle.  We assume that the ions inside the layer are uniformly distributed.  The ions outside the layer are described by the distribution functions $n_s(\vec{x})$, where $s= \pm 1,q$ is the ionic species.  Formally, Debye-Huckel theory can be derived by expanding the free energy to quadratic order in the difference $n_s (\vec{x})-n_s$ \cite{Schiessel}, where $n_s$ is the bulk concentration of ion species $s$.  Assuming the ions are point-like and discarding terms that contribute constants to the free energy,
 \begin{eqnarray}
\label{eq:Fent}
& &\beta F_{ent}=\sum_s\Bigg(\frac{n_s}{2}\int \limits_{r>R} d^3x \left[  \frac{n_s (\vec{x})}{n_s} -1\right]^2 +n_s \alpha_{st} L+\Bigg.\\
\nonumber
& &\Bigg. +\lambda_s^{st} L\left[\ln\left( \frac{\lambda_s^{st}}{n_s \alpha_{st}}\right)-1\right]\Bigg) + \lambda_q  M L\left[\ln\left( \frac{\lambda_q}{n_q\alpha_b}\right)-1\right]
\end{eqnarray}
where $\lambda_q L$ is the number of multivalent ions condensed on each rod in the bundle, $\alpha_b M L =\pi L \left(a^2-a_0^2\right)$ is the volume available to the ions inside the bundle, $\lambda_s^{st} L$ is the number of $s$-valent ions in the Stern layer, and $\alpha_{st} L \approx 2 \pi R w L$ is the volume of the layer ($w\ll R$ is the width of the layer). 

In order to calculate the mean-field electrostatic energy of the system, we model each bundle as a homogeneously charged cylinder of radius $R \approx a \sqrt{M}$ (so that its volume is approximately equal to the bundle volume), and the surrounding Stern layer as a uniform surface charge distribution (i.e. we set $w=0$).  That is, $-e n_b(\vec{x}) = -\theta (R-r) e\lambda/\pi a^2$ and $e n_{st} (\vec{x}) =  \theta (R-r) e \lambda_{st}/2\pi R$ are the charge distributions of a bundle and its Stern layer, respectively, where $-e \lambda \equiv -e \left(\lambda_0-q\lambda_q\right)$ is the renormalized linear charge density of one rod in the bundle and $e\lambda_{st} \equiv e \left( \lambda_1^{st}+q \lambda_q^{st}-\lambda_-^{st}\right)$ is the linear charge density of the Stern layer. The electrostatic free energy is given by
\bb
\label{eq:Fes}
\beta F_{ES}=\frac{l_B}{2} \iint \frac{d^3 x d^3 x'}{| \vec{x}-\vec{x}' |} n_{tot}(\vec{x})n_{tot}(\vec{x}')
\en
where $l_B = e^2/\epsilon k_B T$ is the Bjerrum length ($l_B \approx 7.1$\AA\, in water) and  $e n_{tot}(\vec{x}) = e \sum_s s n_s (\vec{x})\theta (r-R)+e n_{st} (\vec{x})- e n_b (\vec{x})$ is the total charge distribution for a WS cell.  
 
The correlation-induced attractive interactions lead to the formation of a ``bond'' of energy $E_{bond}= - \gamma k_B T L$ between each pair of neighboring rods in the bundle.  If we assume that the rods in the bundle are packed in a hexagonal array, as has been experimentally observed \cite{Bloomfield, Hud, Needleman1, Angelini}, then, to a very good approximation for all $M \geq 2$, $\beta F_{attr} = \beta E_{bond} B =-\gamma L \left( 3 M- 3.6 \sqrt{M} \right)$, where $B$ is the number of bonds in a bundle.  The first term is the bulk attractive energy; the second term is an effective surface tension that arises from the fact that the rods on the bundle surface have less neighbors than the bulk rods.  Thus, we can see that each bundle in our model is equivalent to a homogeneously charged cylinder with an effective surface tension in the presence of counterions.  This is very similar to the Rayleigh instability \cite{Rayleigh} of a charged water droplet in the presence of counterions \cite{Deserno2}.

The number of ions inside each bundle and its surrounding Stern layer, as well as the distribution of ions outside of these regions, minimize the total free energy $\beta F$ (subject to the constraint of overall charge neutrality). For the ion distributions $n_s (\vec{x})$, this minimization yields the expected DH distributions, $n_s(\vec{x}) = n_s\left[1-s\psi(\vec{x})\right]$, where the dimensionless electrostatic potential $\psi (\vec{x})$ is given by
\bb
\psi (r) = 
\begin{cases}
\frac{l_B \lambda}{a^2}\left(r^2-R^2\right)-\frac{2 l_B \lambda_{tot} K_0(\kappa R)}{\kappa R K_1(\kappa R)} & r<R\\
-\frac{2 l_B \lambda_{tot} K_0(\kappa r)}{\kappa R K_1(\kappa R)} & r>R
\end{cases}
\en
Here, $- e \lambda_{tot} = -e \lambda M +e \lambda_{st}$ is the total linear charge density of the bundle and its Stern layer, $\kappa^2 = 4 \pi l_B \sum_s s^2 n_s$, and $K_\nu (x)$ is the modified Bessel function of the second kind of order $\nu$.   Clearly, both $\gamma$ and $\beta F_{corr}$ should depend on $\lambda_q$.  However, since the precise nature of the counterion correlations is not specified in our model, we ignore this dependence; that is, we calculate $\lambda_q$ at the mean-field level.  Discarding all constant terms, the free energy eq. (\ref{eq:freeEnergy}) can be written as $\beta F = N \left[ a^2 \mathcal{F}_1/R^2 + \mathcal{F}_2 L\right]$ where $\mathcal{F}_1$ is given by the final three terms of eq. (\ref{eq:Fent}) and 
\bb
\label{eq:F2}
\mathcal{F}_2=\frac{3.6 \gamma a}{R}+\frac{l_B \lambda^2 R ^2}{4 a^2} +\frac{l_B a^2 \lambda_{tot}^2 K_0 (\kappa R)}{\kappa R^3 K_1 (\kappa R)}
\en
The minimization conditions for $\lambda_q$ and $\lambda_s^{st}$ are, respectively:
\bb
\label{eq:lambdaQ}
\ln \left(\frac{\lambda_q}{n_q \alpha_b}\right)- \frac{l_B q \lambda R^2}{2 a^2}- \frac{2 l_B q \lambda_{tot} K_0 (\kappa R)}{\kappa R K_1 (\kappa R)} =0
\en
\bb
\label{eq:lambdaST}
\ln \left(\frac{\lambda_s^{st}}{n_s \alpha_{st}}\right) - \frac{2 l_B s \lambda_{tot} K_0(\kappa R)}{\kappa R K_1(\kappa R)}=0, \quad s=\pm 1, q
\en
\begin{figure}
\includegraphics[scale=.9]{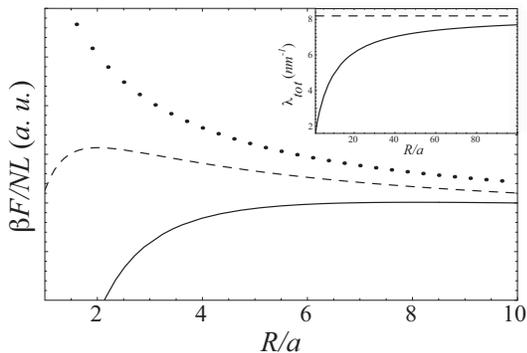}
\caption{\label{fig1}Free energy $\beta F (R)$ \cite{fignote} for $a_0=1\,nm,\,a=1.4\,nm, \, w=1 \, nm,\,  \lambda_0^{-1}=0.17\,nm, \, n_1=10 \, mM,\,q=3, \, \gamma=1.4 \, nm^{-1}$, and $n_q=0.1 \, \mu M$ (solid line), $10 \, \mu M$ (dashed line), and $1 \, mM$ (dotted line). Inset: $\lambda_{tot} (R)$ for the above parameter values with $n_q = 10 \, \mu M$.  The dashed line indicates the asymptotic value of $\lambda_{tot}$ (see text).}
\end{figure}
If we solve eqs. (\ref{eq:lambdaQ}) and (\ref{eq:lambdaST}), we can see that $\lambda_{tot}$ approaches a constant at large $R$ (see inset of Fig. \ref{fig1}).  Indeed, in the limit $R \rightarrow \infty$, the second term in eq. (\ref{eq:lambdaQ}) -- which is due to the electrostatic self-energy of the bundle -- dominates, causing $\lambda \rightarrow \frac{2a^2}{l_B q R^2} \ln\left(\frac{\lambda_0}{q n_q \alpha_b}\right)$ and $\lambda_{tot} \rightarrow \frac{M \lambda}{1+ \kappa w}$.  Thus, as the bundle grows, additional counterions condense inside the bundle, so that the total charge of the bundle remains constant.

The equilibrium bundle radius $R_{eq}$ is given by the value of $R$ that minimizes the free energy.  Fig.  \ref{fig1} shows the free energy as a function of $R$ for various values of $n_q$.  We can see that as soon as the attractive energy is strong enough to induce bundle formation, $R_{eq} \rightarrow \infty$.  This is a result of the large number of counterions that enter each bundle, which causes the entropic and electrostatic resistance to bundle growth to be weak: in the limit $R \rightarrow \infty$, $1/M(\beta F_{ES}+\beta F_{ent}) \sim -1/R^2$.  This resistance is overwhelmed by the attractive energy $\beta F_{attr}/M \sim 1/R$, thus causing the equilibrium bundle size to diverge.  

As stated above, these results do not take into account the dependence of $\gamma$ and $\beta F_{corr}$ on $\lambda_q$.  It is important to note, however, that the asymptotic results $\left| \lambda \right| \sim 1/R^2$ and $\beta F \sim 1/R -1/R^2$ hold for \textit{any} such dependence (even those which cause overcharging): when the bundle is large, the long-range mean-field self-energy of the bundle dominates over the short-range correlation energy, causing the renormalized charge density of the bundle to be small and the resistance to bundle growth to be weak.

\begin{figure}
\includegraphics[scale=.9]{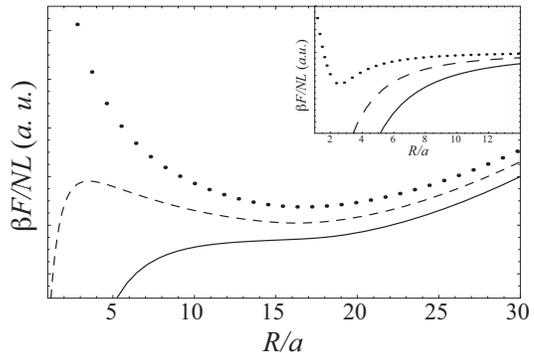}
\caption{\label{fig2} Free energy $\beta F(R)$ \cite{fignote} for finite size ions inside the bundle, with $\gamma = 1.7 \, nm^{-1}$ and $\lambda^* = 0.01 \lambda_0$ (the remaining parameter values are the same as in Fig. \ref{fig1}). Inset: $\beta F(R)$ \cite{fignote} for frustrated attractive interactions, with $\xi=0.1, \,\phi_{max}=2$, and $\gamma = 1.7 \, nm^{-1}$ (the remaining parameter values are the same as in Fig. \ref{fig1}).}
\end{figure}

In the model discussed above, the density of ions inside the bundle can in principle be arbitrarily high.  In reality, however, both steric and short-range electrostatic interactions limit the ion density inside the bundle.  Steric interactions prevent the ion density inside the bundle from exceeding the close packing density.  In addition, our model underestimates the strong repulsion between the ions inside the bundle when their density is high, because the charge of these ions is smeared out over the entire bundle in calculating $\beta F_{ES}$.  This additional short-range electrostatic repulsion effectively increases the size of the ions in the bundle.  If we treat the ions inside the bundle as finite size particles with an effective volume $v_{eff}$, then    $\beta F_{ent} \rightarrow \beta F_{ent} + M L (\Lambda-\lambda_q) \ln(1-\lambda_q/\Lambda)+\lambda_q$, where $\Lambda \approx \alpha_b/v_{eff}$ is the maximum number of ions per unit length that can condense on a single rod.  This adds a term $-\ln(1-\lambda_q/\Lambda)$ to the RHS of eq. (\ref{eq:lambdaQ}) which diverges as $\lambda_q \rightarrow \Lambda$, thus forcing $\lambda_q<\Lambda$ for any bundle size $R$.  If $\lambda^* \equiv \lambda_0-q \Lambda>0$, then the asymptotic result $\lambda \sim 1/R^2$ no longer holds; rather, $\lambda \rightarrow \lambda^*$ for large $R$.  In order for $\lambda^*>0$, the counterions must exceed a minimum effective size set by the spacing between the rods in the bundle (we assume that this spacing is set by the minimum of the short-range correlation-induced attraction energy, which is independent of the bundle size).  For DNA, where $1/\lambda_0=1.7$\AA~, $a_0=10$\AA, and  $a \approx  14$\AA~ in the presence of trivalent cobalt hexammine \cite{Bloomfield,Hud}, this implies the effective radius of the ions $\delta_{eff} \gtrsim 7$\AA. When $\lambda^*>0$, the self-energy of the bundle -- in particular, the second term in eq. (\ref{eq:F2}) -- diverges at large $R$, as shown in Fig. \ref{fig2}.  A local minimum in $\beta F(R)$ is obtained only if the surface tension dominates the free energy for some values of $R$.  As shown in Fig. \ref{fig2}, this is not the case at small multivalent salt concentrations, as the total attraction is not large enough to overcome the electrostatic repulsion between rods.  As $n_q$ increases, however, more ions enter inside the bundle and reduce the electrostatic repulsion for small bundles, thus creating a local minimum in $\beta F(R)$ that is  primarily determined by the balance of the first two terms in eq. (\ref{eq:F2}), $R_{eq} \approx a \left( \frac{2 \gamma}{l_B \lambda^{*2}}\right)^{1/3}$.  Notice that the bundling transitions is discontinuous, and that the bundle size should be independent of further increases in $n_q$, as has been recently observed for microtubule bundles \cite{Needleman2}.

Up to this point, we have assumed that the attraction between neighboring rods is independent of the bundle size. There are several mechanisms, however, that can frustrate the bonds between rods as the bundle grows.  For short-ranged, pairwise additive interactions, only a fraction of ions in a narrow ``contact stripe'' between the two rods become correlated with one another \cite{Shklovskii}.  When rod-rod dimers form, the size of the contact stripe is maximized.  If the size of this stripe is large enough, then the bond energy will decrease as the bundle grows and the rods have to divide the ions in their condensed layers equally among all of their bonds.  Alternatively, a non-uniform PE charge distribution results in a relative orientation that minimizes the electrostatic repulsion between two neighboring rods. Achieving the optimum orientation between a rod and all of its nearest neighbors may cost energy or be physically impossible; in either case, the bond energy effectively decreases as the bundle grows.  Indeed, it has been experimentally observed that F-actin filaments undergo twist distortions when forming bundles to reduce the electrostatic repulsion between neighboring rods \cite{Angelini}.  If we write $\beta F_{attr} = - \gamma L M \phi \left(\frac{B}{M}\right)$, where $B/M \equiv b$ is the average number of bonds per rod ($b<3$ for hexagonally ordered bundles), then $\phi(b)=b$ for unfrustrated interactions.  To encapsulate the effects of bond frustration, we chose $\phi(b)$ to be a hyperbola that approaches the asymptote $\phi=b$ for small $b$ and $\phi=\phi_{max}$ for large b, $\phi(b)=\frac{1}{2}\left(b+\phi_{max}\right)-\frac{1}{2}\left| \phi_{max}-b\right|\sqrt{1+\frac{\xi}{\left(b-\phi_{max}\right)^2}}$.  In other words, the total attractive energy gained for each rod in the bundle saturates as $b$ increases; by adjusting $\phi_{max}$ and $\xi$, we can control the saturating value and rate of saturation, respectively.   As shown in the inset of Fig. \ref{fig2}, a local minimum in $\beta F (R)$ can be obtained by choosing certain values of $\phi_{max}$ and $\xi$.  Unlike the minimum obtained with finite-size ions, this effect cannot lead to arbitrarily large bundles; rather, the onset of frustration must occur at a sufficiently small bundle size, or the entropic and electrostatic resistance to bundle growth will not be large enough to prevent infinite bundles.   

In summary, we have introduced a mean-field model to calculate the equilibrium bundle size of highly charged, rodlike polyelectrolytes in the presence of multivalent counterions.   For point-like counterions and a short-range attraction that is independent of the bundle size, many counterions enter inside the bundle, causing the short-range attraction to overwhelm the resistance to bundle growth and leading to infinite bundles at equilibrium.   In order to obtain a finite equilibrium bundle size, then, either the electrostatic self-energy must be enhanced or the attractive energy suppressed for large bundles.  The former can be accomplished if the short-range interactions between the ions inside the bundle are large enough to prevent the neutralization of large bundles by counterion condensation; the latter, if the interactions between rods in the bundle become frustrated as the bundle grows.  

\begin{acknowledgments}
The authors would like to thank A. Y. Grosberg, C. R. Safinya, D. J. Needleman, and C.D. Santangelo for useful discussions.  We acknowledge the support of the MRL Program of the National Science Foundation under Award No. DMR00-80034 and NSF Grant No. DMR02-037555.   MLH also acknowledges the support of a National Science Foundation Graduate Research Fellowship.  
\end{acknowledgments}


\begin{thebibliography}{}

\bibitem{Sader} J. E. Sader and D. Y. C. Chan, {Langmuir} {\bf 16}, 324 (2000).

\bibitem{Bloomfield} V. A. Bloomfield, {Biopolymers} {\bf 44}, 269 (1997), and references therein.

\bibitem{Hud} N. V. Hud and K. H. Downing, {Proc. Natl. Acad. Sci.} {\bf 98}, 14925 (2001); C. C. Conwell, I. D. Vilfan, and N. V. Hud,  {Proc. Natl. Acad. Sci.} {\bf 100}, 9296 (2003).

\bibitem{Tang1} J. X. Tang, S. Wong, P. T. Tran, and P. A. Janmey, {Ber. Bunsenges. Phys. Chem.} {\bf 100}, 796 (1996).

\bibitem{Angelini} T. E. Angelini, H. Liang, W. Wriggers, and G. C. L. Wong {Proc. Natl. Acad. Sci.} {\bf 100}, 8634 (2003).  

\bibitem{Needleman1} D. J. Needleman, M. A. Ojeda-Lopez, U. Raviv, H. P. Miller, L. Wilson, and C. R. Safinya, {Proc. Natl. Acad. Sci.} {\bf 101}, 16099 (2004).

\bibitem{Tang2} J. X. Tang, P. A. Janmey, A. Lyubartsev, and L. Nordenski\"{o}ld, {Biophys. J.} {\bf 83}, 566 (2002).

\bibitem{Jensen} N. Gr\o nbech-Jensen, R. J. Mashl, R. F. Bruinsma, and W. M. Gelbart, \prl {\bf 78}, 2477 (1997).

\bibitem{Stevens} M. J. Stevens, \prl {\bf 82}, 101 (1999).

\bibitem{Deserno1} M. Deserno, A. Arnold, and C. Holm, {Macromolecules} {\bf 36}, 249 (2003).

\bibitem{Guldbrand} L. Guldbrand, L. G. Nilsson, and L. Nordenski\"{o}ld, {J. Chem. Phys} {\bf 85}, 6686 (1986). 

\bibitem{Nordenskiold} L. Nordensiki\"{o}ld and A. Lyubartsev, {J. Phys. Chem} {\bf 99}, 10373 (1995).

\bibitem{Allahyarov} E. Allahyarov, G. Gompper, and H. L\"{o}wen, \pre {\bf 69}, 041904 (2004). 

\bibitem{Oosawa} F. Oosawa, {Biopolymers} {\bf 6}, 1633 (1968).

\bibitem{Ha1} B. -Y. Ha and A. J. Liu, \prl {\bf 79}, 1289 (1997).

\bibitem{Rouzina} I. Rouzina and V. A. Bloomfield, {J. Phys. Chem.} {\bf 100}, 9977 (1996).

\bibitem{Shklovskii} B. I. Shklovskii, \prl {\bf 82}, 3268 (1999).

\bibitem{delaCruz1} F. J. Solis and  M. O. de la Cruz, \pre {\bf 60} 4496 (1999).

\bibitem{Netz}  A. Naji and R. R. Netz, {Eur. Phys. J. E} {\bf 13}, 43 (2004).

\bibitem{Podgornik} R. Podgornik and V. A. Parsegian, \prl {\bf 80}, 1560 (1998).

\bibitem{Ha2} B. -Y. Ha and A. J. Liu, \prl {\bf 81}, 1011 (1998); {Europhys.  Lett.} {\bf 46}, 624 (1999).

\bibitem{delaCruz2} C. Huang and M. O. de la Cruz, {Macromolecules} {\bf 35}, 976 (2002).

\bibitem{Henle} M. L. Henle and P. A. Pincus, \emph{in preparation}.

\bibitem{Manning} G. S. Manning, {Q. Rev. Biophys.} {\bf 7}, 179 (1969).

\bibitem{Schiessel} M. N. Tamashiro and H. Schiessel, {J. Chem. Phys} {\bf 119}, 1855 (2003); \pre {\bf 68}, 066106 (2003).

\bibitem{Rayleigh} Lord Rayleigh, {Philos. Mag.} {\bf 14}, 184 (1882).

\bibitem{Deserno2} M. Deserno, {Eur. Phys. J. E} {\bf 6}, 163 (2001).

\bibitem{Needleman2} D. J. Needleman, private communication.

\bibitem{fignote} Note that the zero of each free energy curve has been chosen so as to clarify the figures. 

\end{thebibliography}
 \end{document}